\documentclass[iop]{emulateapj}
\usepackage{natbib}
\usepackage{graphicx,color,rotating}
\usepackage{footnote,lineno}
\usepackage{ulem}
\usepackage{xspace}
\usepackage{graphicx,amssymb,amsmath,amsfonts,times,hyperref}

\begin{document}

\onecolumngrid{
\hfill {\tt MCTP-17-20}

\hfill {\tt MIT-CTP/4945}

\hfill {\tt FERMILAB-PUB-17-427-A}

\hfill {\tt PUPT-2538}

\vskip 0.2in

\title{Comment on ``Characterizing the population of pulsars in the Galactic bulge with the
  {\it Fermi} Large Area Telescope'' [arXiv:1705.00009\MakeLowercase{v}1]}

\author{
Richard Bartels,\altaffilmark{1}
Dan~Hooper,\altaffilmark{2,3,4}
Tim~Linden,\altaffilmark{5}
Siddharth~Mishra-Sharma,\altaffilmark{6}
Nicholas~L.~Rodd,\altaffilmark{7}
Benjamin~R.~Safdi,\altaffilmark{8}
Tracy~R.~Slatyer\altaffilmark{7}
}

\altaffiltext{1}{GRAPPA and Institute of Physics, University of Amsterdam, Science Park 904, 1098XH Amsterdam, The Netherlands}
\altaffiltext{2}{Center for Particle Astrophysics, Fermi National Accelerator Laboratory, Batavia, IL 60510}
\altaffiltext{3}{Department of Astronomy and Astrophysics, University of Chicago, Chicago, IL 60637}
\altaffiltext{4}{Kavli Institute for Cosmological Physics, University of Chicago, Chicago, IL 60637}
\altaffiltext{5}{Center for Cosmology and AstroParticle Physics (CCAPP), Ohio State University, Columbus, OH 43210}
\altaffiltext{6}{Department of Physics, Princeton University, Princeton, NJ 08544}
\altaffiltext{7}{Center for Theoretical Physics, Massachusetts Institute of Technology, Cambridge, MA 02139}
\altaffiltext{8}{Leinweber Center for Theoretical Physics, Department of Physics, University of Michigan, Ann Arbor, MI 48109}
\setcounter{footnote}{0}

\begin{abstract}
The {\it Fermi}-LAT Collaboration recently presented a new catalog of gamma-ray sources located within the $40^{\circ}$$\times$~$40^{\circ}$ region around the Galactic Center~\citep{Fermi-LAT:2017yoi} -- the Second Fermi Inner Galaxy (2FIG) catalog. Utilizing this catalog, they analyzed models for the spatial distribution and luminosity function of sources with a pulsar-like gamma-ray spectrum.
\cite{Fermi-LAT:2017yoi} v1 also claimed to detect, in addition to a disk-like population of pulsar-like sources, an approximately 7$\sigma$ preference for an additional centrally concentrated population of pulsar-like sources, which they referred to as a ``Galactic Bulge'' population.
Such a population would be of great interest, as it would support a pulsar interpretation of the gamma-ray excess that has long been observed in this region. In an effort to further explore the implications of this new source catalog, we attempted to reproduce the results presented by the {\it Fermi}-LAT Collaboration, but failed to do so. Mimicking as closely as possible the analysis techniques undertaken in \cite{Fermi-LAT:2017yoi}, we instead find that our likelihood analysis favors a very different spatial distribution and luminosity function for these sources. Most notably, our results do not exhibit a strong preference for a ``Galactic Bulge" population of pulsars. Furthermore, we find that masking the regions immediately surrounding each of the 2FIG pulsar candidates does {\it not} significantly impact the spectrum or intensity of the Galactic Center gamma-ray excess. Although these results refute the claim of strong evidence for a centrally concentrated pulsar population presented in \cite{Fermi-LAT:2017yoi}, they neither rule out nor provide support for the possibility that the Galactic Center excess is generated by a population of low-luminosity and currently largely unobserved pulsars.
In a spirit of maximal openness and transparency, we have made our analysis code available at {\url{https://github.com/bsafdi/GCE-2FIG}}.

\end{abstract}

\maketitle

\section{A Comparison with Ajello {\it et al.}}
\setcounter{footnote}{0}

The {\it Fermi}-LAT Collaboration recently presented the Second Fermi Inner Galaxy (2FIG) source catalog~\citep{Fermi-LAT:2017yoi}.\footnote{A revised version of \cite{Fermi-LAT:2017yoi} is being submitted simultaneously with (and in response to) this comment. Throughout this work, when we discuss \cite{Fermi-LAT:2017yoi}, we refer to the original version arXiv:1705.00009v1.} This catalog consists of 374 sources that have been detected with a test statistic (TS) of 25 or greater, located within the $40^{\circ}\times 40^{\circ}$ region surrounding the Galactic Center. Among this list, there are 104 sources (86 of which are not contained in the 3FGL catalog~\citep{Acero:2015hja}) that exhibit best-fit spectral parameters that are characterized as pulsar-like by \cite{Fermi-LAT:2017yoi}.\footnote{Throughout this paper, we will discuss the results presented by \cite{Fermi-LAT:2017yoi} obtained using their ``official'' interstellar emission model (IEM). Although they also present results for an ``alternative'' IEM, they only provide an efficiency function corresponding to the case of the official IEM, making it impossible for us to evaluate the results obtained using the alternative IEM. In any case, the results presented by the {\it Fermi}-LAT Collaboration are nearly identical regardless of which IEM model was adopted.} More specifically, \cite{Fermi-LAT:2017yoi} classifies a source as a pulsar candidate if its spectrum prefers a power-law with an exponential cutoff over that of a simple power-law at a level of TS~$>$~9 and is best-fit by a spectral index $\Gamma < 2$ and a cutoff energy $E_{\rm cut} <10$ GeV.

\begin{table*}
\center
\begin{tabular}{cccccc}
\multicolumn{6}{c}{Results of~\cite{Fermi-LAT:2017yoi}}  \\ 
\hline
\hline
  $N_{D}$        & $z_0$[kpc]       & $\beta$ &  $N_{B}$ &  $\alpha$  & TS  \\
\hline
 $22500^{+5200}_{-4800}$	&	$0.71^{+0.16}_{-0.16}$ &	$1.34^{+0.07}_{-0.07}$ & 0 &  \nodata &  0  \\
 $3560^{+980}_{-870}$	&	$0.72^{+0.17}_{-0.17}$ &	$1.24^{+0.06}_{-0.06}$ & $1330^{+270}_{-210}$ &  $2.60$ &  63   \\
  $3610^{+1010}_{-930}$ &	$0.75^{+0.18}_{-0.18}$ &	$1.25^{+0.07}_{-0.07}$ & $1370^{+280}_{-220}$ &  $2.57^{+0.23}_{-0.23}$ &  69  \\
  \hline
  \hline
  \\ 
 \multicolumn{6}{c}{Results of This Study} \\ 
  \hline
  \hline
  $N_{D}$        & $z_0$[kpc]       & $\beta$ &  $N_{B}$ &  $\alpha$  & TS  \\
\hline
     $(1.26^{+0.48}_{-0.40})$$\times$$10^6$	&	$0.13^{+0.06}_{-0.04}$ &	$2.08^{+0.07}_{-0.07}$ & 0 &  \nodata &  0 \\
$(1.06^{+0.42}_{-0.34})$$\times$$10^6$	&	$0.08^{+0.05}_{-0.03}$ &	$2.11^{+0.08}_{-0.07}$ & $(5.03^{+4.89}_{-2.52})$$\times$$10^5$ &  $2.60$ &  8.3 \\
  $(1.04^{+0.40}_{-0.34})$$\times$$10^6$ &	$0.09^{+0.05}_{-0.03}$ &	$2.11^{+0.07}_{-0.07}$ & $(8.30^{+11.50}_{-5.16})$$\times$$10^5$ &  $2.78^{+0.15}_{-0.34}$ & 8.5 \\
  \hline
  \hline
\end{tabular}
\caption{The best-fit values and $1\sigma$ uncertainty for the number of disk pulsars, $N_{D}$, the scale-height of the disk population, $z_0$, the index of the luminosity function, $\beta$, the number of bulge pulsars, $N_{B}$, and the slope of the bulge population's inner profile, $\alpha$. Also listed is the value of the test statistic (TS) with respect to the disk-only hypothesis (first row). In the second and third rows, results are shown with the inclusion of a bulge-like component, fixing the profile of that component to $\alpha=2.6$ or letting $\alpha$ float, respectively. All of the results shown here have utilized the ``official'' interstellar emission model (as presented by \cite{Fermi-LAT:2017yoi}) and were calculated using $3.3^{\circ}$ spatial bins. The results of this study (bottom) vary substantially in almost every respect from those found by~\cite{Fermi-LAT:2017yoi} (top).}
\label{tab:logresults}
\end{table*}

By combining the Galactic coordinates and fluxes of these sources with an efficiency function that describes the probability of detecting a given source at a particular sky location and flux, one can test various models for the underlying spatial distribution and luminosity function of the pulsar-like source population. For the disk-like component of pulsars, \cite{Fermi-LAT:2017yoi} adopt the standard Lorimer distribution~\citep{Lorimer:2003qc}:
\begin{equation}
n_{\rm disk} \propto R^n \, \exp(-R/\sigma) \, \exp(-|z|/z_0)\,,
\end{equation}
with $n=2.35$ and $\sigma=1.528$ kpc. The quantities $R$ and $z$ represent the location of the source in cylindrical coordinates. The vertical scale height of this distribution, $z_0$, is allowed to float in the fit. 

In addition to this disk population of pulsars, \cite{Fermi-LAT:2017yoi} include a centrally concentrated and spherically symmetric population, described as follows:
\begin{equation}
n_{\rm bulge} \propto r^{-\alpha}, \,\,\, \,\,\, r<3 \, {\rm kpc}\,,
\end{equation}
where $r$ is the distance from the Galactic Center. We will refer to this centrally located source population as the ``bulge'' population. The parameter $\alpha$ is either set to 2.6 in order to match the spatial distribution of the observed gamma-ray excess~\citep{Goodenough:2009gk, Abazajian:2012pn, Gordon:2013vta, Daylan:2014rsa, Calore:2014xka, TheFermi-LAT:2015kwa, TheFermi-LAT:2017vmf} or is allowed to float.

For the gamma-ray luminosity function of these sources, \cite{Fermi-LAT:2017yoi} adopt a power-law functional form, $dN/dL \propto L^{-\beta}$, which is assumed to extend unbroken between $10^{31}$ and $10^{36}$ erg/s (integrated from 0.3 GeV to 500 GeV). 

For a given spatial distribution and luminosity function, the expected number of sources in a given spatial bin (labeled by $i$, $j$) and flux bin (labeled by $k$) is calculated as follows:
\begin{eqnarray}
N_{i,j,k}^{\rm model} &=& \omega_{i,j,k} \int_0^\infty ds \, s^2 \int_{\Delta \Omega_{i,j}} d\Omega \, \int^{4\pi s^2 S^{\rm max}_k}_{4 \pi s^2 S^{\rm min}_k}  \frac{dN}{dL} dL \nonumber \\
&\times&\big[n_{\rm disk}(s,l,b) + n_{\rm bulge}(s,l,b) \big]  \,,
\end{eqnarray}
where $s$ denotes the distance along the line-of-sight, $l$ and $b$ are Galactic coordinates, $S_k^{\rm min}$ and $S_k^{\rm max}$ correspond to the range of fluxes across bin $k$ (integrated between 0.3 and 500 GeV), and $\omega_{i,j,k}$ is the efficiency factor, defined as the probability that a point source with a pulsar-like spectrum present in a given spatial and flux bin will be detected and included in the 2FIG catalog. Following \cite{Fermi-LAT:2017yoi}, we adopt a 12$\times$12 array of equally sized spatial bins across the $40^{\circ}$$\times$$40^{\circ}$ region-of-interest and 8 logarithmically-spaced flux bins, with six equally sized bins spanning the range of  (1--10)~$\times$~10$^{-6}$~MeV~cm$^{-2}$~s$^{-1}$ and two larger logarithmically spaced bins covering the range of (1--10)~$\times$~10$^{-5}$~MeV~cm$^{-2}$~s$^{-1}$.

The model prediction for the expected source distribution can then be compared to the pulsar candidates in the inner galaxy, binned identically and labelled $N_{i,j,k}^{\rm obs}$. The fitting is performed using a Poisson inspired likelihood:\footnote{There is a typo in the sign of the last two terms of this equation in \cite{Fermi-LAT:2017yoi}. We have confirmed with the corresponding authors of that work that the signs are correct in the underlying code, and this is not the origin of our disagreement.}
\begin{eqnarray}
\ln \mathcal{L} = \sum_{i,j,k} \left[ N_{i,j,k}^{\rm obs} \ln \left( N_{i,j,k}^{\rm model} \right) - N_{i,j,k}^{\rm model} \right] - \mathcal{L}_{\rm prior}\,.
\end{eqnarray}
The final term here, $\mathcal{L}_{\rm prior}$, was applied by \cite{Fermi-LAT:2017yoi} in order to ensure that the number of very bright pulsars predicted by their model is in reasonable agreement with the number of pulsars observed across the sky. More specifically, they apply the following Gaussian prior:
\begin{equation}
\mathcal{L}_{\rm prior} = \frac{(N^{\rm model}_{S>S_0}(\lambda)-N^{\rm data}_{S>S_0})^2}{2\sigma_N^2}\,,
\label{prior}
\end{equation}
where $S_0 \equiv 1.8\times 10^{-5}$ MeV cm$^{-2}$ s$^{-1}$, $N^{\rm data}_{S>S_0} =174$ and $\sigma_N=63$.\footnote{Although this was not stated in the text of \cite{Fermi-LAT:2017yoi}, we treated $N^{\rm model}_{S>S_0}$ to be the number of sources predicted by the model with a flux between $S_0$ and that of the brightest pulsar in the 3FGL catalog, $3.896 \times 10^{-3}$ MeV cm$^{-2}$ s$^{-1}$. The exact choice of the maximum flux has essentially no impact on our results. }

\cite{Fermi-LAT:2017yoi} present the results of their likelihood analysis for both a disk-only model and for a model which includes both a disk and bulge population of pulsars. Their results are listed in Table~\ref{tab:logresults} (top), compared with those found by our likelihood analysis of the same list of sources (bottom), which we now describe in detail.

In order to replicate the analysis of \cite{Fermi-LAT:2017yoi}, we performed a parameter scan using \texttt{MultiNest}, which efficiently implements nested sampling of the posterior distribution in the Bayesian framework~\citep{Feroz:2008xx,Buchner:2014nha}. We performed the fit with the \texttt{nlive} parameter set to 1500, specifying the number of live points used during the posterior sampling. In Figs.~\ref{fig:nd}, \ref{fig:ndnb} and \ref{fig:ndnbalpha}, we present the results of our likelihood analysis, for each of the three cases listed in Table~\ref{tab:logresults}.  Those figures display the posterior distribution, with 2-dimensional 1, 2, and 3$\sigma$ contours around the best-fit points indicated.  The dotted vertical lines indicate the 16, 50, and 84$^\text{th}$ percentiles for the 1-dimensional posteriors, while the solid green lines indicate the best-fit values found in \cite{Fermi-LAT:2017yoi}.   In producing these results, we adopted flat priors for each free parameter, extending from $N_{\rm B}=(0$-$3)\times 10^7$, $N_{\rm D}=(0$-$3)\times 10^7$, $\alpha=2.1$-$5.0$, $z_0=0.01$-2.0 kpc, and $\beta=1.1$-$3.0$. In order to maximize the transparency of these results, we have made all of the programs and data inputs/outputs utilized in this analysis publicly available.\footnote{\url{https://github.com/bsafdi/GCE-2FIG/}}

\begin{figure}
	\centering
\includegraphics[width=1.03\columnwidth]{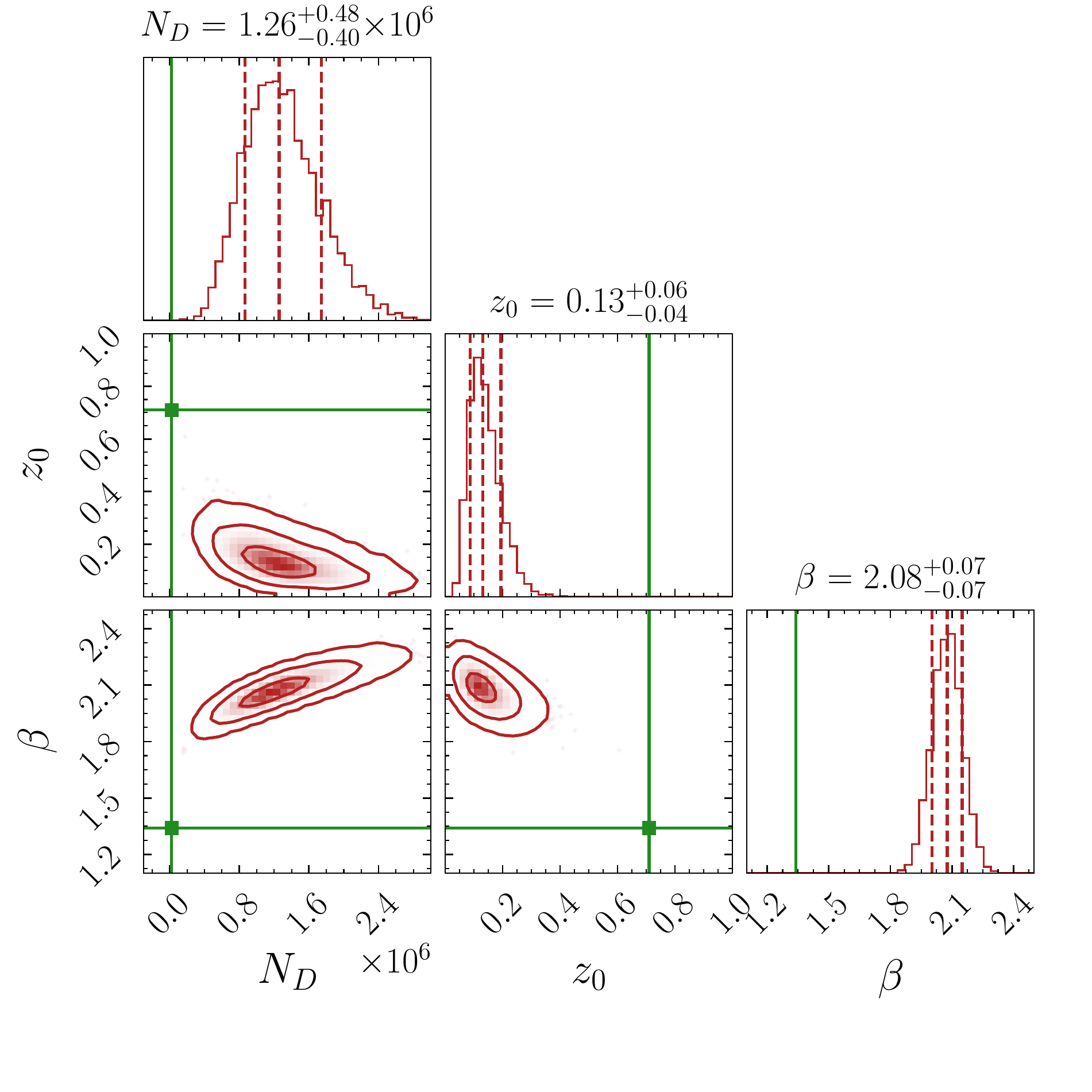}
\vspace{-0.6cm}
\caption{The range of parameters favored by our fit when no bulge-like component is present (corresponding to the first row in Table~\ref{tab:logresults}).  The two-dimensional contours indicate the 1, 2, and 3$\sigma$ 2-D contours around the best-fit points, while the vertical dotted lines in the 1-dimensional posteriors indicate the 16, 50, and 84$^\text{th}$ percentiles.  The solid green lines mark the best-fit values found in \cite{Fermi-LAT:2017yoi} and displayed in the first row of Table~\ref{tab:logresults}.}
\label{fig:nd} 
\end{figure}

\begin{figure}
	\centering
\includegraphics[width=1.03\columnwidth]{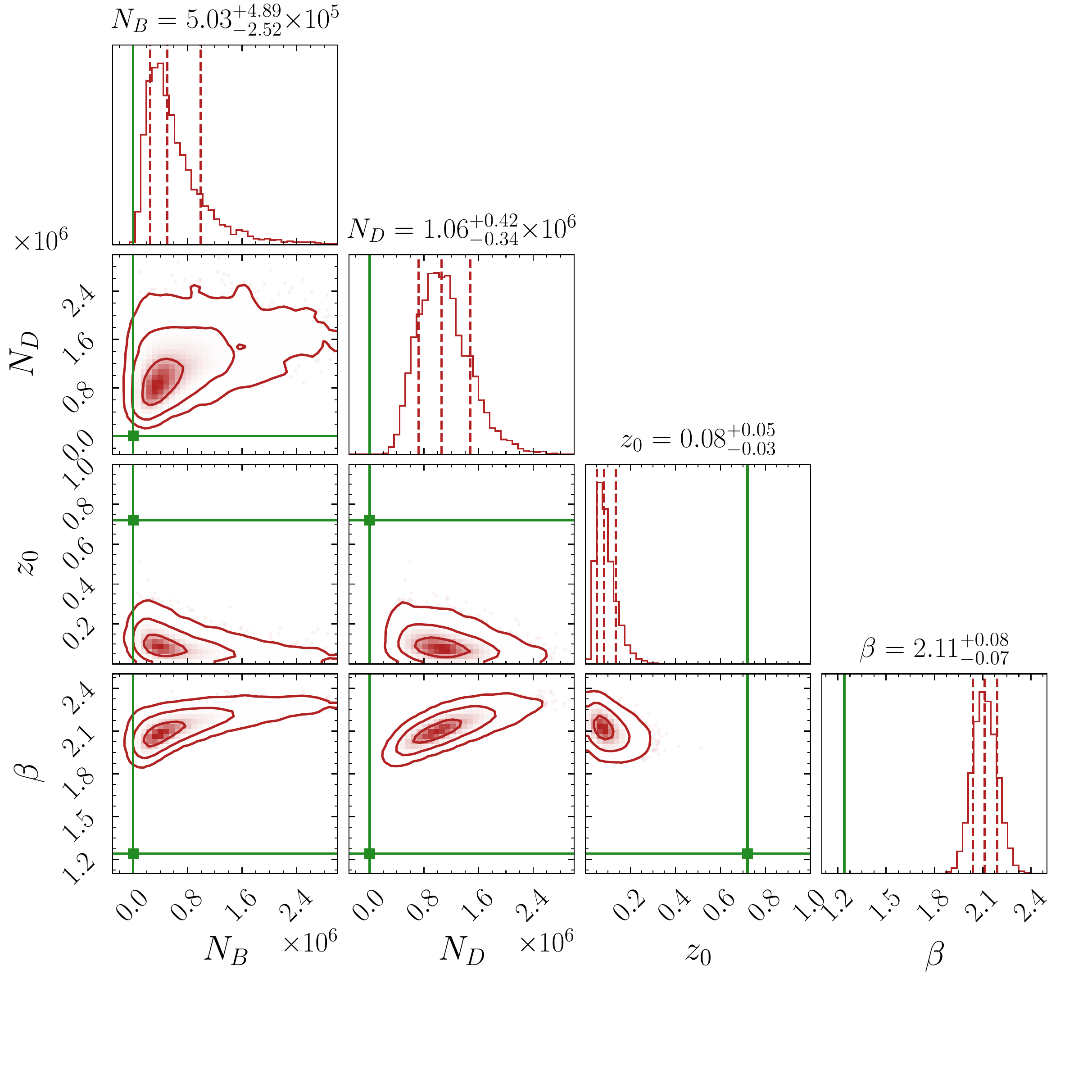}
\vspace{-0.9cm}
\caption{The range of parameters favored by our fit when a bulge-like component with a $\alpha=2.6$ profile is allowed to be present (corresponding to the second row in Table~\ref{tab:logresults}). }
\label{fig:ndnb} 
\end{figure}

\begin{figure*}
	\centering
\includegraphics[width=1.53\columnwidth]{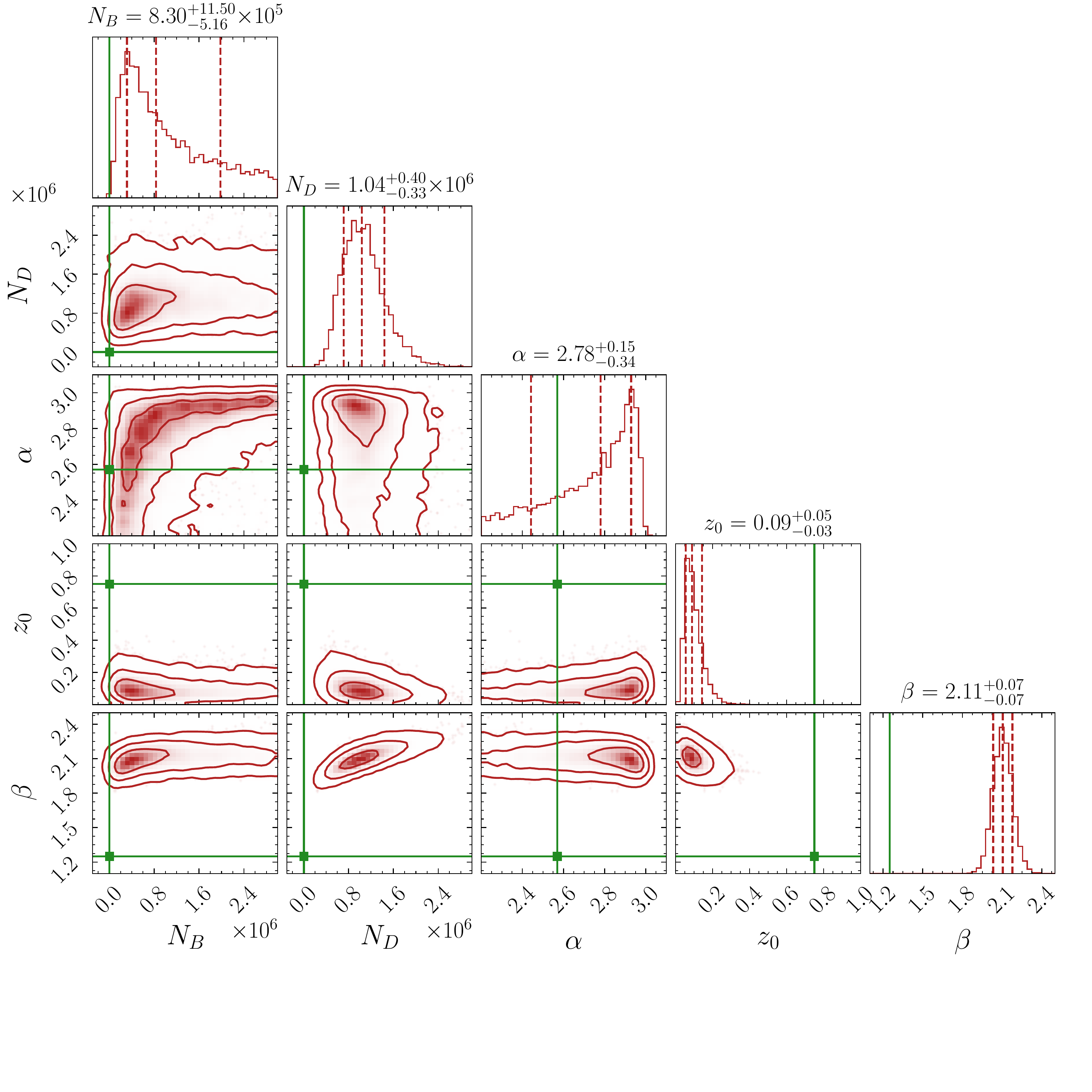}
\vspace{-1.2cm}
\caption{The range of parameters favored by our fit when a bulge-like component with a floating value for the profile slope, $\alpha$, is allowed to be present (corresponding to the third row in Table~\ref{tab:logresults}).}
\label{fig:ndnbalpha} 
\end{figure*}

The results calculated in this work vary substantially in almost every respect from those found by \cite{Fermi-LAT:2017yoi}. In particular, we find that the distribution of sources with pulsar-like spectra are best-fit by a distribution which features an extremely thin disk, $z_0 \sim 0.1$ kpc, and a much softer luminosity function, $\beta \approx 2.1$, than that presented by \cite{Fermi-LAT:2017yoi}. Furthermore, our fit much more modestly prefers the presence of a central source component, at a level of TS~$\sim$\,8, in contrast to TS~$\sim$~63-69 found by \cite{Fermi-LAT:2017yoi}. We additionally note that whereas the application of the prior (see Eq.~\ref{prior}) strongly impacted the results of \cite{Fermi-LAT:2017yoi},\footnote{Mattia Di Mauro, private communication.} our fit naturally yields a total number of bright sources that is compatible to that observed, and the prior has a negligible impact on our results. We will return to discuss the physical significance of these results later in this article.

Throughout this analysis, we attempted to mimic as closely as possible the likelihood calculation presented by \cite{Fermi-LAT:2017yoi}. In particular, we adopted the same bin sizes and identically masked the innermost $2^{\circ}$ in order to avoid problems associated with source confusion. In one respect, however, our analysis necessarily differs from that of \cite{Fermi-LAT:2017yoi}. In particular, \cite{Fermi-LAT:2017yoi} have applied an extended efficiency function that depends on Galactic longitude and accounts for the dispersion between true and observed flux of sources, whereas only a 2-dimensional version of this function, describing the dependence of the efficiency on Galactic latitude and the flux without accounting for dispersion, has been made available to those who are not members of the {\it Fermi}-LAT Collaboration. The publicly available efficiency function corresponds to that shown in Fig.~7 of \cite{Fermi-LAT:2017yoi}. We have been assured by the corresponding authors of this study, however, that the original results of \cite{Fermi-LAT:2017yoi} are only mildly sensitive to the distinction between the true and measured flux of these sources. In an effort to estimate the impact of any longitude dependence in the efficiency function, we have evaluated the efficiency function in each longitude bin after rescaling the flux proportionally to the sensitivity map for pulsar-like sources, as presented by the {\it Fermi}-LAT Collaboration as part of the Second Pulsar Catalog (see Fig.~16 of~\cite{TheFermi-LAT:2013ssa}). The best-fit parameter values we obtained using this modified efficiency function varied by less than 1$\sigma$ from those presented in Table~\ref{tab:logresults} and only modestly increased the preference for a bulge-component, by $\Delta$TS $\sim$ 3.

\section{Dependence on the Parameterization of the Luminosity Function}

We would like to emphasize that although we are confident in our results as presented here, we do not necessarily find their physical significance to be straightforward to interpret. More specifically, we do not necessarily believe that the best-fit parameters of our study reflect an accurate description of the distribution of pulsars in the Milky Way. For one thing, while we have utilized the \cite{Fermi-LAT:2017yoi} classification of 2FIG sources as ``pulsar-like" or ``blazar-like", we are not necessarily convinced that all, or even most, of the sources within the 2FIG catalog that are classified as ``pulsar-like", are, in fact, pulsars. Although \cite{Fermi-LAT:2017yoi} demonstrate that the spectral shapes of pulsars and blazars can be used among sources in the 3FGL catalog to efficiently differentiate these source classes, it seems likely
that such techniques will be far less effective for the much fainter, and thus much less well measured, sources that dominate the 2FIG catalog.  The efficiency functions we used, taken from \cite{Fermi-LAT:2017yoi}, were also calibrated for a source population resembling the best-fit model of that work, which differs substantially from ours; the efficiency functions for our best-fit model might be non-negligibly modified due to the different distribution of sources by latitude and flux.\footnote{We thank the corresponding authors of \cite{Fermi-LAT:2017yoi} for clarifying the methodology utilized in their efficiency function.}

Furthermore, the very narrow disk distribution favored by our fit ($z_0 \sim 0.1$ kpc) appears inconsistent with the combined young and millisecond pulsar population
as identified by radio observations, which is $\sim 0.3$ kpc. A thin disk with a scale height of $\sim 0.1$ kpc is appropriate for young pulsars, however, millisecond pulsars are
expected to follow a thicker distribution with a scale height of $\sim 0.5$ kpc and contribute significantly to the overall pulsar population
 \citep{2004A&A...425.1009M, Lorimer:2006qs, Levin:2013usa, Calore:2014oga}.
We consider it likely that this inconsistency is in large part the result of our luminosity function parameterization, and we note that past studies have found there to be significantly fewer low-luminosity pulsars than this power-law parameterization would suggest. In particular, the luminosity function of millisecond pulsars can be well-fit by either a broken power-law or log-parabola, centered around intermediate luminosities (roughly 10$^{33}$~erg~s$^{-1}$)~\citep{Cholis:2014noa,Hooper:2015jlu,Hooper:2016rap}. The fact that low-luminosity sources are so prevalent in our best-fit model forces the majority of the pulsar contribution to stem from very local sources, significantly impacting the value of the disk width, $z_0$, preferred by our fit.


\begin{table*}
\center
\begin{tabular}{cccccc}
\multicolumn{6}{c}{$L_{\rm min}=10^{32}$ erg/s} \\ 
\hline
\hline
  $N_{D}$        & $z_0$[kpc]       & $\beta$ &  $N_{B}$ &  $\alpha$  & TS\\
\hline
$(1.21^{+0.44}_{-0.35})$$\times$$10^5$	&	$0.19^{+0.07}_{-0.05}$ &	$2.08^{+0.10}_{-0.09}$ & 0 &     \nodata &  0 \\
 $(1.07^{+0.45}_{-0.33})$$\times$$10^5$	&	$0.13^{+0.06}_{-0.04}$ &	$2.15^{+0.12}_{-0.10}$ & $(5.14^{+5.50}_{-2.62})$$\times$$10^5$ &  $2.60$ &  8.1 \\  \hline \hline
\\ 
\multicolumn{6}{c}{$L_{\rm min}=10^{33}$ erg/s} \\
\hline 
\hline
  $N_{D}$        & $z_0$[kpc]       & $\beta$ &  $N_{B}$ &  $\alpha$  & TS\\
\hline
$(1.24^{+0.36}_{-0.29})$$\times$$10^4$	&	$0.32^{+0.08}_{-0.06}$ &	$2.10^{+0.13}_{-0.13}$ & 0 &  \nodata &  0 \\
$(1.02^{+0.40}_{-0.29})$$\times$$10^4$	&	$0.23^{+0.09}_{-0.06}$ &	$2.20^{+0.17}_{-0.14}$ & $(4.57^{+3.95}_{-2.07})$$\times$$10^3$ &  $2.6$ &  10.1 \\
\hline
\hline
\end{tabular}
\caption{As in Table~\ref{tab:logresults}, but adopting a minimum luminosity of $10^{32}$ erg/s (top) or $10^{33}$ erg/s (bottom). By increasing the value of $L_{\rm min}$ relative to that adopted by \cite{Fermi-LAT:2017yoi}, we find that our fit can accommodate values of $z_0$ which are more consistent with the results of radio surveys.}
\label{tab:logresultsLmin}
\end{table*}

To explore the impact of the lowest luminosity pulsars in our fit, we show in Table~\ref{tab:logresultsLmin} results adopting minimum luminosities of $L_{\rm min}=10^{32}$ and $10^{33}$ erg/s (in contrast, \cite{Fermi-LAT:2017yoi} adopted $L_{\rm min} =10^{31}$ erg/s). As anticipated, this change results in significantly larger values of $z_0$, in greater concordance with radio observations. We also note that this modification does not significantly alter the degree to which the fit prefers the presence of a bulge population of pulsars.


\begin{figure}[t]
\centering
\includegraphics[width=0.9\columnwidth]{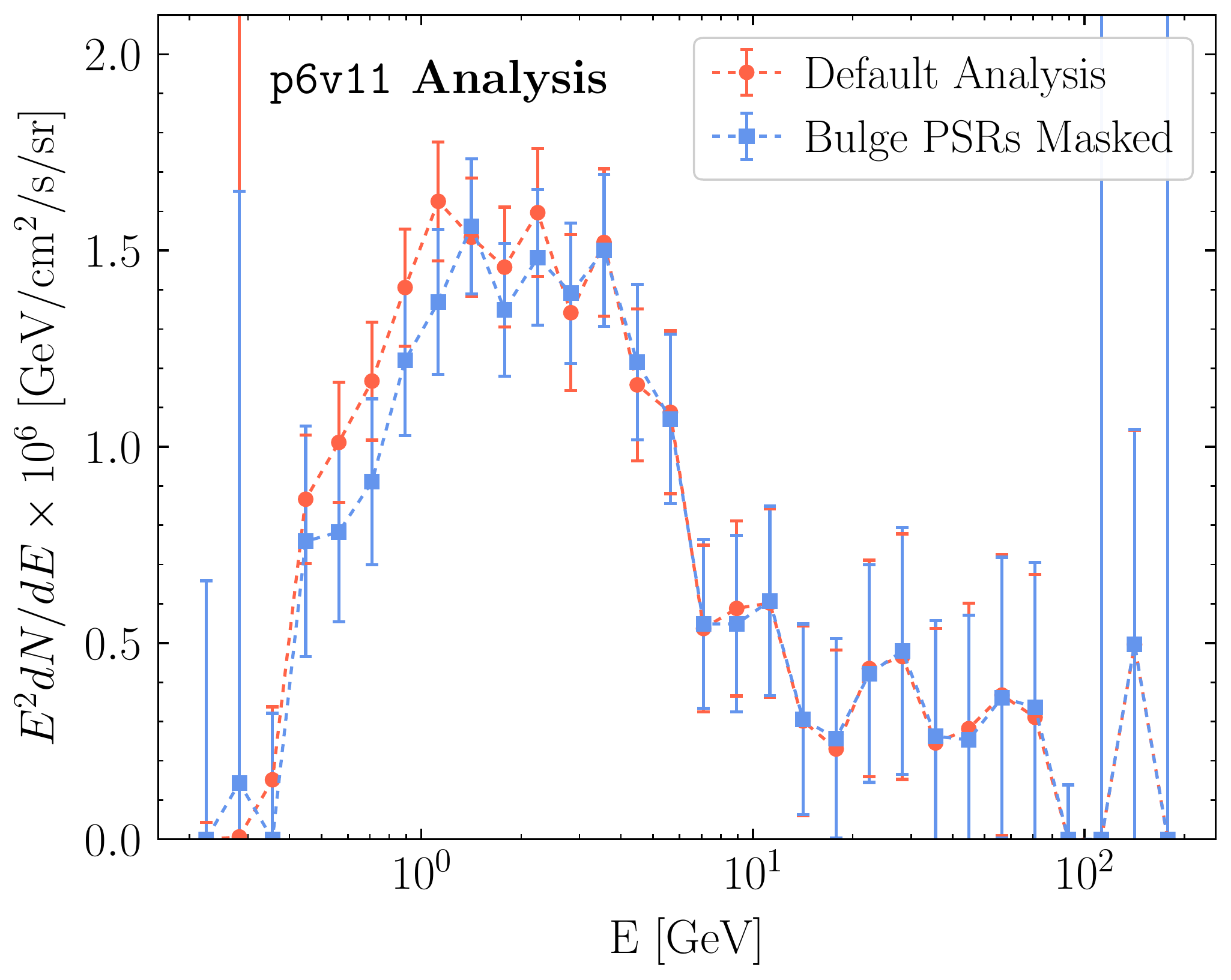} \\
\includegraphics[width=0.9\columnwidth]{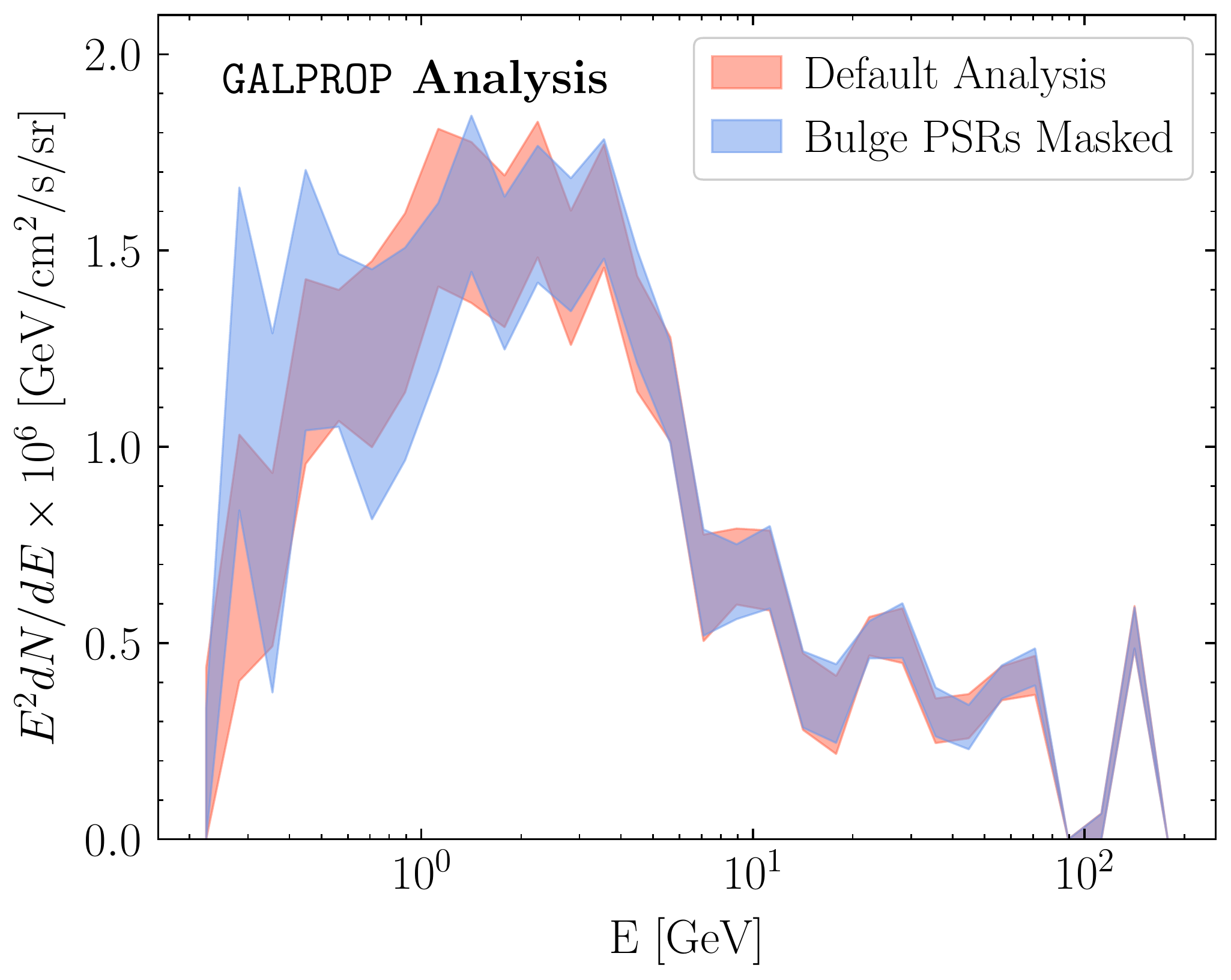}
\caption{The spectrum and intensity of the Galactic Center gamma-ray excess performed over the standard region-of-interest (red), and after masking the 95\% containment regions surrounding each of the 86 pulsar candidates contained within the 2FIG catalog and identified using the ``official'' diffuse emission model, as well as all identified 3FGL pulsars (blue). In the upper frame, we show the best-fit values and statistical error bars, as found using the \texttt{p6v11} diffuse emission model. In the lower frame, the bands represent the envelope of best-fit values found over an ensemble of 14 \texttt{GALPROP} models (without error bars). From this comparison, it is clear that masking these sources has a negligible impact on the intensity and spectral shape of the measured excess.}
\label{fig:GCEtempFit}
\end{figure}

\section{The Robustness of the Galactic Center Gamma-Ray Excess To The Masking of 2FIG Pulsar Candidates}

If there exists a centrally-concentrated population of pulsars with the characteristics claimed by \cite{Fermi-LAT:2017yoi}, 
we should expect the brightest of these sources to account for the majority of the gamma-ray emission associated with this population.
In particular, their best-fit population can account for all Galactic-Center excess emission \citep[see Fig.~5 of ][]{Fermi-LAT:2017yoi}, however, 
given the efficiency function we expect $\sim 70\%$ of the emission to be resolved as pulsar candidates. Consequently, the 2FIG should absorb
most of the excess.

To test this hypothesis, we have carried out a basic template analysis, similar to that performed in~\cite{Daylan:2014rsa}, with and without masking the 95\% containment radii regions around each of the 86 pulsar candidate sources contained in the 2FIG catalog and identified using the ``official'' diffuse model, as well as all 3FGL pulsar candidates in the region.\footnote{Specifically those sources classified as either \texttt{PSR} or \texttt{psr} in the 3FGL catalog.} In the case that the brightest 2FIG pulsars provide a significant fraction of the flux associated with the Galactic Center excess, this analysis should find a significantly diminished excess component. The results of this test are shown in Fig.~\ref{fig:GCEtempFit}. From this comparison, it is clear that masking these sources has a negligible impact on the intensity and spectral shape of the measured excess.

In producing Fig.~\ref{fig:GCEtempFit}, we utilized the top quartile of UltracleanVeto Pass 8 events collected between August 4, 2008 and July 7, 2016. We also applied the following quality cuts: zenith angle greater than $90^{\circ}$ and \texttt{(DATA\_QUAL$>$0)\&\&(LAT\_CONFIG==1)}.\footnote{For a complete list of recommended data criteria, see \url{https://fermi.gsfc.nasa.gov/ssc/data/analysis/documentation/Cicerone/Cicerone_Data_Exploration/Data_preparation.html}.} This set of gamma rays is then binned spatially into an \texttt{nside}=256 \texttt{HEALPix} map~\citep{Gorski:2004by} and into 30 logarithmically spaced energy bins between 200 MeV and 200 GeV. We adopted a $30^{\circ}$$\times$$30^{\circ}$ region-of-interest around the Galactic Center, masking the regions within $1^{\circ}$ of the Galactic Plane and around the 68\% containment radii of the 300 brightest and most variable sources contained in the 3FGL catalog~\citep{Acero:2015hja}.\footnote{Our region-of-interest is somewhat smaller than the $40^{\circ} \times 40^{\circ}$ considered in earlier works, such as~\cite{Daylan:2014rsa} and~\cite{Calore:2014xka}, but was shown in~\cite{Linden:2016rcf} to be more stable for analyzing the excess.} To obtain the results shown in this figure, we performed a template based analysis of this dataset, implemented using \texttt{NPTFit}~\citep{Mishra-Sharma:2016gis}. The fit includes templates intended to describe the Galactic diffuse emission, isotropic flux, emission associated with the \textit{Fermi} bubbles~\citep{Su:2010qj}, emission from the known 3FGL sources, and the flux corresponding to the excess. The morphology of the excess is characterized by a generalized NFW profile squared and integrated along the line-of-sight, adopting an inner slope of $\gamma=1.2$ (equivalent to $\alpha=2.4$), which is the best-fit value for this dataset~\citep{Daylan:2014rsa, Calore:2014xka,Keeley:2017fbz}.

In each of the two frames of Fig.~\ref{fig:GCEtempFit}, the results shown in red (blue) are those corresponding to the analysis performed without (with) a mask for the 2FIG pulsar candidates and identified 3FGL pulsars. In the upper frame, we adopted the \texttt{p6v11} \textit{Fermi} diffuse model.\footnote{This is a common abbreviation for the full name of this model, which is gll\_iem\_v02\_P6\_V11\_DIFFUSE. The model is available here: \url{https://fermi.gsfc.nasa.gov/ssc/data/p6v11/access/lat/BackgroundModels.html}.} Although this is not the latest diffuse model released by the {\it Fermi}-LAT Collaboration, the more recent models have had large-scale residuals added back in, such as those associated with the \textit{Fermi} bubbles or even the excess itself. For this reason, the most recent diffuse models are unsuitable for studying the properties of the Galactic Center excess. In the lower panel, we show the envelope of the best-fit spectra that is found across a range of 14 \texttt{GALPROP} based diffuse emission models.\footnote{These models are taken from~\cite{Calore:2014xka}. Adopting their nomenclature, these are referred to as models A and F-R (F-R originally appeared in~\cite{Ackermann:2012pya}).} For each of these models, we used separate templates to describe the emission associated with firstly the $\pi^0$ and bremsstrahlung emission and secondly the inverse Compton emission.



\section{Summary and Discussion}

In this study, we have attempted to utilize the sources contained in the Second Fermi Inner Galaxy (2FIG) catalog to characterize the spatial distribution and luminosity function of those sources which exhibit a pulsar-like gamma-ray spectrum. In doing so, we attempted to mimic the analysis techniques employed by the {\it Fermi}-LAT Collaboration, but found that our likelihood analysis favors a very different spatial distribution and luminosity function for these sources. Most notably, our results do not exhibit a strong preference for a ``Galactic Bulge" population of pulsars. Whereas \cite{Fermi-LAT:2017yoi} find strong evidence (TS~$\sim 60-70$) in support of a centrally concentrated population of pulsar-like gamma-ray sources, we find a significantly weaker preference for any such population (TS~$\sim 8$).

Furthermore, we find that masking the regions immediately surrounding each of the 2FIG pulsar candidates does not significantly impact the spectrum or intensity of the Galactic Center gamma-ray excess. We thus conclude that the pulsar candidates contained in the 2FIG catalog do not substantially contribute to the observed excess, in contrast to what is implied by the best-fit luminosity function of \cite{Fermi-LAT:2017yoi}.

We would like to emphasize that we are {\it not} attempting to make the case here that there is not a significant population of pulsars located in the Inner Galaxy, or that such sources are not potentially responsible for the Galactic Center gamma-ray excess. Instead, we have argued that the characteristics of the 2FIG catalog, as presented by \cite{Fermi-LAT:2017yoi}, do not provide significant support for the existence of such a source population. Past studies have identified evidence of small scale power in the gamma-ray emission from the Inner Galaxy~\citep{Lee:2015fea,Bartels:2015aea}, suggestive of an unresolved point source population. On the other hand, the relative lack of both bright pulsars~\citep{Hooper:2016rap,Hooper:2015jlu,Cholis:2014lta} and low-mass X-ray binaries~\citep{Haggard:2017lyq,Cholis:2014lta} in the Inner Galaxy suggests that if such a pulsar population is in fact responsible for the excess emission, that population would have to exhibit rather different characteristics than those observed in the disk of the Milky Way and in globular clusters. Regardless of these and other arguments, the question of the origin of the Galactic Center gamma-ray excess remains an open one. Our analysis of the characteristics of the 2FIG catalog does not provide significant support for either a pulsar or a dark matter interpretation of this signal.


\bigskip

{\it Note added: While completing the final stages of this manuscript, we were in regular contact with the corresponding authors of \cite{Fermi-LAT:2017yoi}, and provided them with our results and code. While comparing their results with those presented here, the authors of \cite{Fermi-LAT:2017yoi} identified, and alerted us to, an error in their analysis framework. When corrected, their analysis yields results that are more consonant with the analysis shown here. The revised version of \cite{Fermi-LAT:2017yoi}, to appear simultaneously with this work, removes the incorrect analysis.}

\vspace{0.5cm}

\begin{acknowledgements}

We would like to sincerely thank the corresponding authors of \cite{Fermi-LAT:2017yoi} for taking the time to clarify and explain many aspects of the {\it Fermi}-LAT Collaboration's analysis \citep{Fermi-LAT:2017yoi}. 
R.~Bartels is supported by the Netherlands Organization for Scientific Research (NWO) through a GRAPPA-PhD fellowship. TL acknowledges support from
NSF Grant PHY-1404311 to John Beacom. T.~R.~Slatyer and N.~L.~Rodd are supported by the U.S. Department of Energy under grant Contract Numbers DE$-$SC00012567 and DE$-$SC0013999. This manuscript has been authored by Fermi Research Alliance, LLC under Contract No. DE-AC02-07CH11359 with the U.S. Department of Energy, Office of Science, Office of High Energy Physics. The United States Government retains and the publisher, by accepting the article for publication, acknowledges that the United States Government retains a non-exclusive, paid-up, irrevocable, world-wide license to publish or reproduce the published form of this manuscript, or allow others to do so, for United States Government purposes. This research made use of the NASA Astrophysics Data System (ADS) and the IDL Astronomy User's Library at Goddard.

\end{acknowledgements}

\bibliography{2figcomment}
\end{document}